\documentclass[usenatbib]{mn2e}

\usepackage{graphicx,tabularx,amsmath,amssymb,upgreek,wasysym,mathtools,multirow}
\numberwithin{equation}{section}
\title[RCW 120: A possible case of hit and run]{RCW 120: A possible case of hit and run, elucidated by multitemperature dust mapping}
\author[K. A. Marsh \& A. P. Whitworth]
{K. A. Marsh$^{1,2}$\thanks{E-mail: kam@ipac.caltech.edu}
and A. P. Whitworth$^{1}$ \\ \\
$^{1}$School of Physics and Astronomy, Cardiff University, Cardiff CF24 3AA, Wales, UK\\
$^{2}$Infrared Processing and Analysis Center, California Institute of Technology, Pasadena, CA 91125, USA}
\begin{document}
\maketitle
\label{firstpage}

\begin{abstract} 
We present resolution-enhanced images of warm dust at multiple temperatures 
and opacity index values in the star-forming bubble/HII region, RCW 120. 
The image set, representing a 4D hypercube of differential column 
density, was obtained using our Bayesian procedure, {\sc ppmap}. 
The cool peripheral material ($\sim16$--22 K) exhibits ragged clumpy  
structure as noted previously by others. However, at 
higher temperatures ($\stackrel{>}{_\sim}26$ K)
the geometry changes dramatically, showing a bubble boundary which is
accurately circular in projection, except for the previously-reported
opening in the north. Comparison with {\it Spitzer\/} 8 $\mu$m data
suggests that the $\sim26$--30 K dust seen by {\it Herschel\/} resides in
the photodissociation region (PDR) surrounding the HII region. Its
projected radial profile is consistent with that of a spherical 
shell, thus arguing against previous suggestions of cylindrical or planar 
geometry. The inferred geometry is, in fact, consistent with previous 
interpretations of the HII region as a classical Str\"omgren sphere, except
for the fact that the ionising star (CD -38$^\circ$\!\!.11636; O8V) is 
displaced by more than half a radius from its geometric centre. 
None of the previously published models has satisfactorily
accounted for that displacement. It could, however, be explained 
by proper motion of the O star at $\sim2$--4 km s$^{-1}$ since its
formation, possibly due to a cloud-cloud collision.
We suggest that the current spherical bubble constitutes the fossilised 
remnant of the initial expansion of the HII region following the formation of
the star, which now continues to flee its formation site.
\end{abstract}

\begin{keywords}
ISM: individual objects (RCW 120) -- ISM: bubbles -- {\it (ISM:)\/} HII regions -- {\it (ISM:)\/} dust, extinction -- stars: formation -- stars: early-type 
\end{keywords}

\section{Introduction}

RCW 120 is a Galactic HII region enclosed by a bubble of gas and dust
which appears as a highly symmetrical shell in {\it Herschel\/} images. This
shell is thought to be the site of triggered star formation by the collect
and collapse process \citep{zav07} or radiatively-driven implosion of
pre-existing condensations \citep{walch15}. The origin of the shell,
however, is not clear. Two possibilities that have been proposed are:
\begin{enumerate}
\item The {\it expanding HII region\/} model. In this interpretation, an
O star forms within an approximately uniform interstellar medium (ISM),
giving rise to a surrounding Str\"omgren sphere of ionised gas. The
expansion of this gas results in a shock front which travels outwards,
initially at the
sound speed, sweeping out a dense shell of molecular gas \citep{zav07,deharv09}.
A variant of this model involves an O star in a
Bonnor-Ebert sphere rather than a uniform ISM, whereby the O star location
is postulated to be far off-centre \citep{och14}. 
\item The {\it cloud-cloud collision\/} model. Star formation
by cloud-cloud collisions was investigated by \citet{habe92} and proposed
as an explanation for the formation of RCW 120 by \citet{tor15}. Their
model involves a collision velocity of $\sim30$ km s$^{-1}$, as suggested by 
the observed radial velocities of $-8\,{\rm km\,s^{-1}}$ for the ring,
and $-28\,{\rm km\,s^{-1}}$ for an adjacent cloud which appears to
be physically connected.
\end{enumerate}

In the present paper we present the results of our investigation of 
the dust distribution in RCW 120 using
{\it Herschel\/} data in conjunction with
a new Bayesian analysis procedure, {\sc ppmap} 
\citep{mar15} which
yields significantly higher spatial resolution than conventional techniques
and provides information on the distribution of dust temperatures and
dust opacity index values along the line of sight.

\section{Methodology}

In contrast with previous approaches to the dust mapping problem,
the {\sc ppmap} procedure drops the assumptions of uniform 
dust temperature and opacity index along the line of sight
\citep{mar15}\footnote{The original version treated the opacity index,
$\beta$, as constant. We have extended the algorithm by
treating $\beta$ as an additional state variable, as discussed
in detail by \citet{mar18}.}.
After taking proper account of the point-spread function and spectral response 
of the telescope, it returns a four-dimensional hypercube of estimated
differential column density as a function of sky location (RA, Dec), dust
temperature, $T_{_{\rm D}}$ and opacity index, $\beta$, assuming a power-law
variation of opacity with wavelength. It also returns a corresponding
hypercube of uncertainty values.

The procedure assumes only that the 
the emission is optically thin over the range of observed wavelengths. 
It works by defining a grid of discrete values of $x$, $y$, $T_{_{\rm D}}$ 
and $\beta$, which represent sampling locations in the continuous
state space of $(x,y,T_{_{\rm D}},\beta)$. 
It then iterates towards the set of differential column densities, in the
vicinities of those grid points, that best reproduces the observed 
monochromatic intensities.
The resolution on the plane of the sky $(x,y)$ is increased by a factor 
of about 4.5 relative to the standard analysis procedure, yielding maps of
RCW 120 with an angular resolution of $\sim8''$, sampled by $4''$ square pixels.

The dust temperature sampling interval is arbitrary. For this work, we have 
used twelve representative dust temperatures, logarithmically spaced between
8 K and 50 K. 
The opacity index sampling interval is also arbitrary, and we have used 
five representative values equally spaced linearly between 1.5 and 2.5.
Our column density
scale is based on an assumed opacity of 0.1 cm$^2$ g$^{-1}$ at a wavelength
of 300 $\mu$m. This reference opacity is
defined with respect to total mass (dust plus gas). Although observationally
determined, it is consistent with a gas to dust ratio of 100 \citep{hil83}.

\section{Observations}

RCW 120 was observed as part of the Hi-GAL survey \citep{mol2010}
using the {\it Herschel\/} PACS and SPIRE instruments 
which provided continuum images in bands centred on
wavelengths of 70, 160, 250, 350, and 500 $\mu$m. 
The spatial resolution values of these data, i.e., the beam sizes at full
width half maximum (FWHM), are approximately
8.5, 13.5, 18.2, 24.9, and 36.3 arcsec, respectively.
Details of the calibration and map-making procedures are
given by \citet{elia13}.  For our analysis we use PSFs
based on the measured {\it Herschel\/} beam profiles \citep{pog10,griffin13}. 

\section{Results}

The above data have been used to generate a 4D hypercube of differential
column density as a function of sky location, dust temperature, and
dust opacity index, from which we can derive various line-of-sight
integrated quantities such as the
integrated column density of dust plus gas, $N$, the mean 
dust temperature, ${\bar T}$, and the mean dust 
opacity index ${\bar \beta}$. 

\begin{figure}
\hspace*{-0.3cm}\includegraphics[width=90mm]{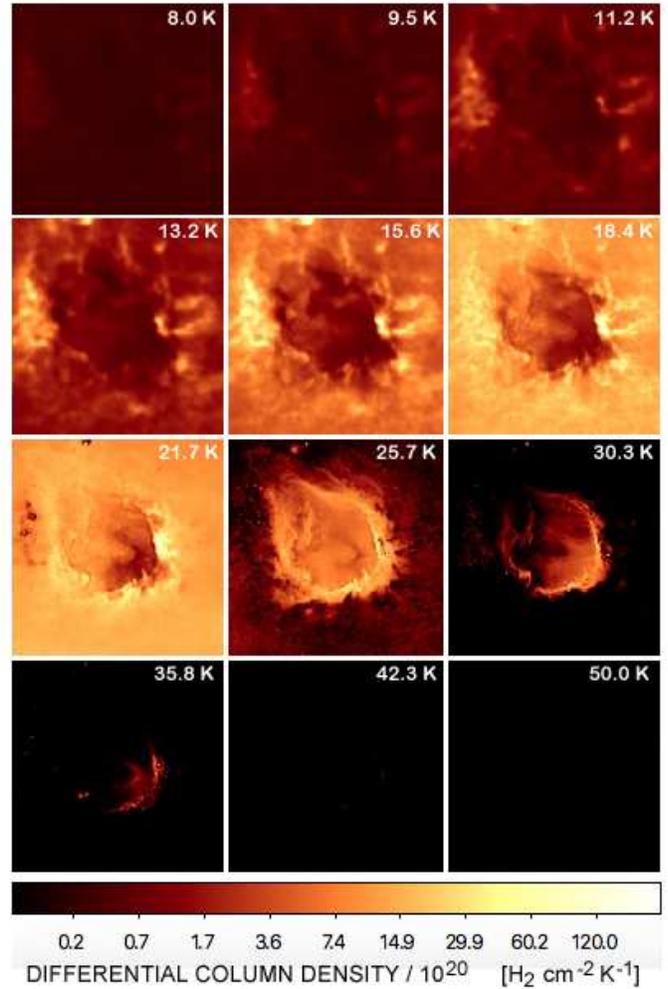}
\caption{Maps of differential column density as a function of temperature.
The field of view in each case is 21 arcmin (8.2 pc)
square.}
\label{fig1}
\end{figure}

\begin{figure}
\hspace*{-0.3cm}\includegraphics[width=90mm]{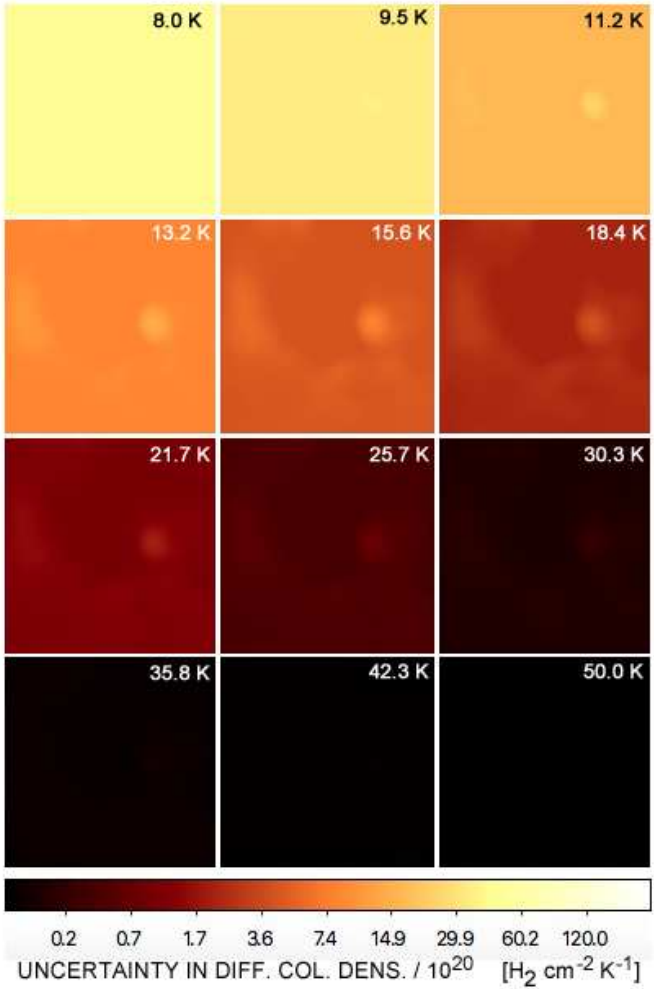}
\caption{Uncertainties in the differential column density maps in Fig. \ref{fig1} as a function of temperature.}
\label{fig2}
\end{figure}

Fig. \ref{fig1} shows a set of images of differential column density
for all 12 temperatures, each image having been summed over opacity index.
The corresponding uncertainties are presented in Fig. \ref{fig2}.

\begin{figure}
\hspace*{-0.3cm}\includegraphics[width=90mm]{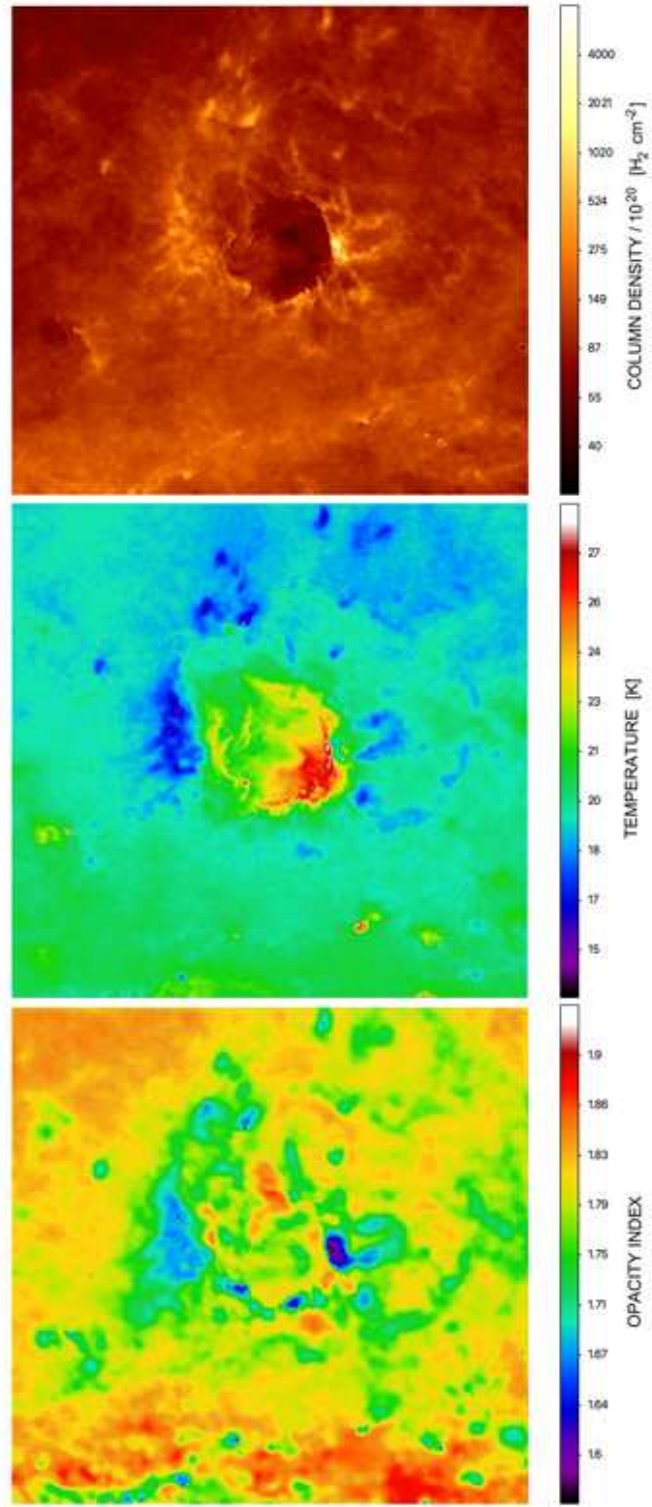}
\caption{Maps of line-of-sight integrated quantities, namely {\it Top:\/} total 
column density, {\it Middle:\/} density-weighted mean dust temperature, and 
{\it Bottom:\/} density-weighted mean dust opacity index. The field of view 
in each case is $42\times42$ arcmin.}
\label{fig3}
\end{figure}

Fig. \ref{fig3} shows the line-of-sight integrated quantities derived from
the differential column density hypercube, namely the total column
density and density-weighted mean values of dust temperature and opacity
index.

\begin{figure}
\includegraphics[width=85mm]{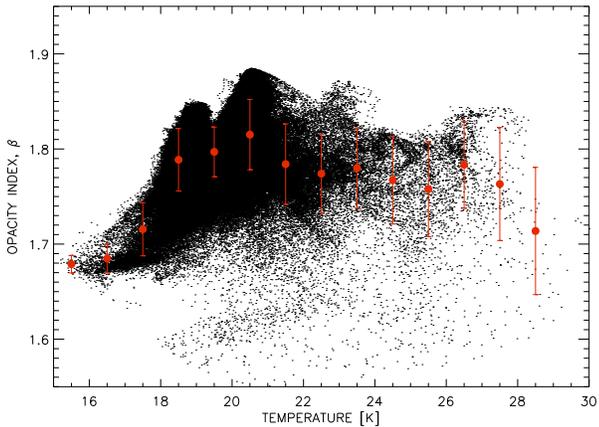}
\caption{Scatter plot of dust opacity index verus temperature. Each black
plotted point represents one pixel from Fig. \ref{fig3}. The red filled
circles and correpsonding error bars represent $2\sigma$ trimmed means and
standard deviations.}
\label{fig4}
\end{figure}

Comparison of the lower two panels in Fig. \ref{fig3} suggests that localised
minima in dust temperature (presumably cores) correspond to local minima
in dust opacity index, $\beta$. This behaviour is quite apparent in the 
scatter plot of $\beta$ versus $T$ shown in Fig. \ref{fig4}, which does
not show the anticorrelation found by \citet{and10}. But as pointed out
by those authors, their anticorrelation may reflect the well-known
$\beta$-$T$ degeneracy in the presence of noise \citep{shet09}. 
Given that the dust is optically thin at all of the
{\it Herschel\/} wavelengths, the occurrence in cool dense clumps,
of $\beta$ values smaller than those in the surrounding ISM,
is suggestive of grain growth \citep{rod06}.

\begin{figure}
\includegraphics[width=82mm]{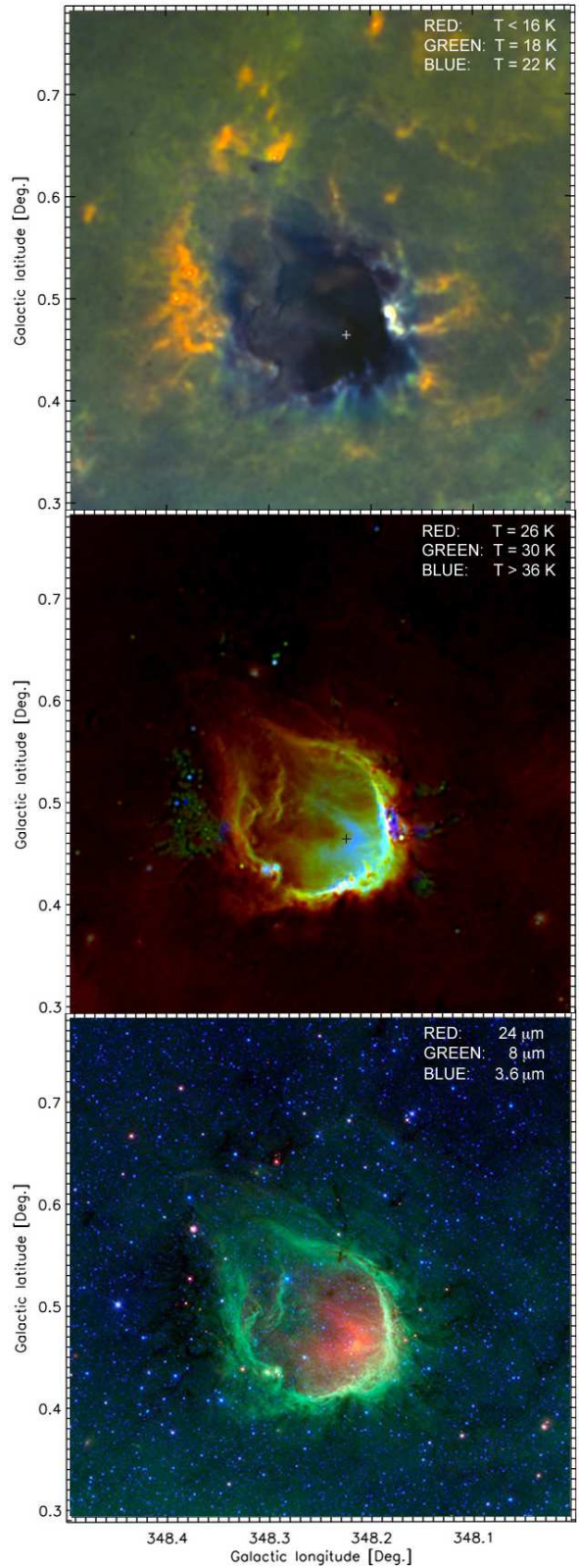}
\caption{Composite images of RCW 120. The top two panels represent
multi-temperature overlays, with red, green, and blue corresponding to
the indicated temperature ranges. The ``+" signs mark the O star position.
For comparison, the bottom panel shows a
multi-wavelength composite from {\it Spitzer\/}. Note the detailed 
correspondence between the 8 $\mu$m IRAC image (shown in green) with
the 26--30 K dust (shown red and green in the middle panel).}
\label{fig5}
\end{figure}

An alternative way of representing the variation of differential column 
density with temperature, besides the multi-panel format of Fig. \ref{fig1},
is to use composite overlays in which different temperatures are represented
by different colours. The upper two panels of Fig. \ref{fig5} present such
multi-temperature composites. In order to cover the entire 8--50 K range
involved in the {\sc ppmap} output, however, it was necessary to split
the temperatures into two separate temperature ranges,
namely (a) cool (8--22 K) and (b) warm (26--50 K).

It is evident that the geometry of RCW 120 is dramatically different when
viewed in the two temperature ranges. At the cooler temperatures
($\stackrel{<}{_\sim}22$ K), a
ragged geometry is evident, corresponding to the presence massive clumps around
the periphery, consistent with the findings of previous authors
\citep{zav07,deharv09}. However, at the
higher temperatures ($\stackrel{>}{_\sim}26$ K)
the geometry changes dramatically, showing a bubble boundary which appears
accurately circular in projection, except for the previously-reported
opening in the north \citep{zav07}. A comparison between the bottom two
panels of Fig. \ref{fig5} shows a detailed correspondence between the 
distribution of 26--30 K dust and that of 8 $\mu$m emission seen by 
{\it Spitzer\/}. This is quite remarkable given that the emission mechanisms
are different in the two cases, representing thermal emission by dust
grains in the former case, and the fluorescence of PAHs in the
ultraviolet field of the ionising star in the case of the 8 $\mu$m emission
\citep{zav07}.  The detailed morphological correspondence
suggests that the $\sim26$--30 K dust seen by {\it Herschel\/} 
resides in the hot photodissociation region (PDR) surrounding the HII region. 

\begin{figure}
\hspace*{-0.3cm}\includegraphics[width=90mm]{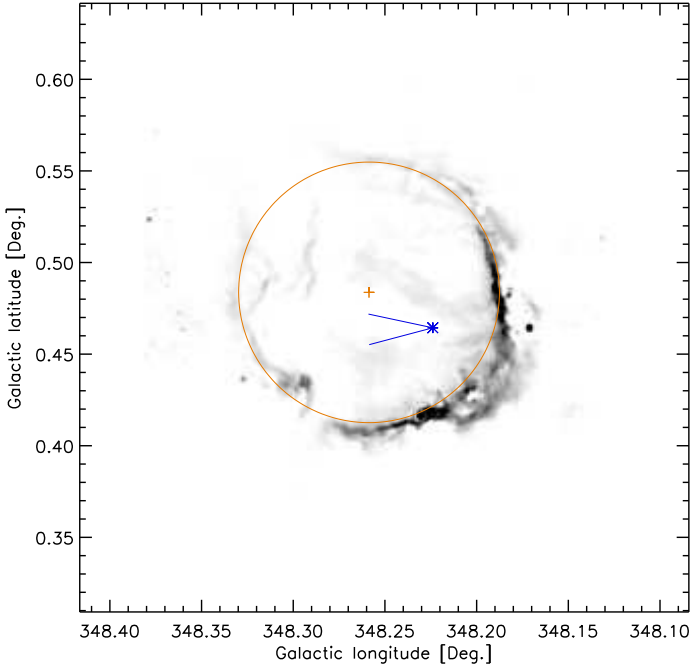}
\caption{The greyscale image represents the bright rim of the 30 K dust 
distribution in RCW 120 which
presumably delineates the edge of the photodissociation region, as discussed
in the text. The best-fit circle is indicated in orange; its geometric centre
is indicated by the ``+" sign. The blue asterisk represents the current
position of the ionising star. The blue lines represent the $1\sigma$
limits of previous proper motion of the O star over the past
$3.2\times10^4$ yr based on {\it Gaia\/} observations. Note, however,
that during this time the ring structure itself would have moved also.}
\label{fig6}
\end{figure}

The geometry of the rim of the bubble is indicated more clearly in
Fig. \ref{fig6} which shows the distribution of differential column density in
a narrow temperature range centred on 30 K after subtracting
a median-filtered background. For comparison, the best-fitting circle
is shown in orange, with the geometric centre indicated by the ``+" sign.
Evidently, the projected edge of the bubble is accurately circular
over a wide angular range ($\sim315^\circ$), and this places constraints
on formation models, as discussed below.

\begin{figure}
\hspace*{-0.3cm}\includegraphics[width=90mm]{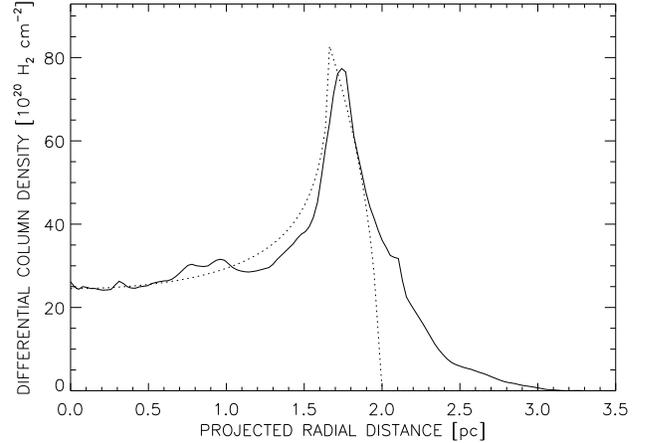}
\caption{The radial distribution of warm dust in the photodissociation region. 
The solid line represents the azimuthally averaged distribution of
differential column density in the temperature range 26--30 K, as estimated
using {\sc ppmap}. It is expressed as a function of the projected distance
from the centre of symmetry as indicated in Fig. \ref{fig6}. The dotted line
represents the theoretical profile based on a 3-parameter model of a hollow 
sphere whose inner radius, outer radius, and central column density are
derived from the observations.}
\label{fig7}
\end{figure}

The question arises as to whether the circular outline represents a
sphere seen in projection. This question is relevant since the 
three-dimensional geometry of such objects such as RCW 120 is still unclear 
\citep{and15}. For example, \citet{deharv09} asks whether the geometry might
involve a preferred plane, while \citet{pav13} suggests that the HII region of 
RCW 120 is cylindrical rather than spherical and that we are observing the
object along the axis of the cylinder. A similar suggestion has been made
by \citet{tor15} in connection with their cloud-cloud collision model.
We can make use of the {\sc ppmap}\ results to investigate this question,
based on the assumption that the 26--30 K dust maps the PDR as suggested
above. Specifically we can check to see if the edge-to-centre contrast
of the differential column density is consistent with that expected for
a hollow shell whose wall has a finite thickness. To facilitate this
check, Fig. \ref{fig7} shows the radial profile of differential column
density, azimuthally averaged about the centre, indicated by
the ``+" sign in Fig. \ref{fig6}. The abcissa of this plot represents
the projected distance, $h$, from the geometric centre. We can compare
this plot with the expected profile, $N_{\rm mod}(h)$, of an optically thin, 
uniform hollow shell with inner and outer radii $r_{\rm in}$ and 
$r_{\rm out}$, respectively, i.e.,
\begin{equation}
N_{\rm mod}(h) = \begin{cases} 2n_0(\sqrt{r_{\rm out}^2-h^2}-\sqrt{r_{\rm in}^2-h^2}) & \mbox{if } h\leq r_{\rm in}, \\
2n_0\sqrt{r_{\rm out}^2 - h^2} & \mbox {if } r_{\rm in} < h < r_{\rm out}, \\ 
0 & \mbox{otherwise, } \end{cases}
\label{eq1}
\end{equation}
where $n_0$ is the volume density, given by $n_0=N_0/(r_{\rm out}-r_{\rm in})$,
and $N_0$ represents the column density through the shell wall at
the projected centre.

We constrain the parameters $N_0$, $r_{\rm in}$, and 
$r_{\rm out}$ using estimates derived from
the observed profile ($N_0=12.25\pm0.07\times10^{20}$ H$_2$ cm $^{-2}$,
$r_{\rm in}=1.66$ pc, $r_{\rm out}=1.98$ pc) and thereby obtain the
model profile shown as a dotted line in Fig. \ref{fig7}. It reproduces
well the behaviour of the observed profile except for the existence of a tail
beyond the peak, which we interpret as the halo of filamentary material
attributed to leakage from the bubble \citep{zav07} and which is
evident in the middle panel of Fig. \ref{fig5}.
In particular, the simple model reproduces
the observed edge-to-centre contrast. The theoretical value for the model
sphere is 3.37, which compares well with the observed value of
$3.15\pm0.18$. This provides support for the interpretation
of spherical geometry for the bubble.

It does, however, contrast sharply with the findings of \citet{pav13} who 
concluded that
a spherical shell does not fit the observed profile of {\it Herschel\/} 
100 $\mu$m emission. However, the latter is not a reliable 
proxy for column density because a temperature gradient is present, such
that the dust temperature decreases outwards 
from the hot ($\sim 30$ K) inner rim. Consequently, the 100 $\mu$m emission is 
biased towards the rim even though there is cooler dust ($\sim 26$ K) further 
out in the PDR, as evidenced by its correspondence with
8 $\mu$m emission as shown in Fig. \ref{fig5}. Thus the PDR shell 
is thicker than suggested by the 100 $\mu$m emission and, as a consequence,
has significant column density through the projected centre of the ring
(see Fig. \ref{fig7}). That feature represents a
key difference from the profile upon which the
conclusions of \citet{pav13} were based.

\section{Discussion}

The above results strongly suggest that the RCW 120 bubble is spherical
except for an opening in the north, attributed to a ``champagne flow" of ionised
gas leaking out of the bubble \citep{zav07}. One particularly 
puzzling aspect, however, is that the 
ionising star, CD$-38^\circ\!\!.11636$ \citep{zav07}, is displaced far from 
the geometric centre of the sphere, as shown in Fig. \ref{fig6}.

The observed geometry is, in fact, inconsistent with all of the formation
models proposed in the literature so far.
In particular, the classical ``expanding HII region" model would place the
ionising star at the centre of the bubble, in contrast to the observed
geometry illustrated in Fig. \ref{fig6}. Although \citet{zav07}
explain the displacement as due to a north-south density gradient,
the ionisation front would then be egg-shaped rather than spherical.
While the displacement could,
in principle, be explained if the star were postulated to be
located off-centre in a Bonnor-Ebert sphere \citep{och14},
such a model leads to an irregularly-shaped ionisation zone, contrary
to observation. In addition, it fails to
reproduce the observed distribution of CO \citep{tor15}. 

In the case of the cloud-cloud collision model as interpreted by
\citet{tor15}, the bubble boundary would correspond to the inner edge of
a U-shaped cavity rather than a near-perfect ring which, again, is
contrary to observation. Although the authors do
point out that a ring-like appearance could result from a viewing angle
along the axis of cylindrical symmetry in their model, the 
star would then appear near the centre of the ring in projection and
would not show the observed displacement.

One piece of information that has not been exploited in previous analyses, 
however, is that the O star has a large proper motion. The {\it Gaia\/} DR2
catalogue indicates motion in RA and Dec of 
$-3.05\pm0.89\,\,{\rm mas}\,{\rm yr}^{-1}$ and 
$-3.12\pm0.63\,\,{\rm mas}\,{\rm yr}^{-1}$, corresponding to a net transverse
motion of $28\pm5\,\,{\rm km}\,{\rm s}^{-1}$, mostly in the direction of 
decreasing Galactic longitude, i.e., away from the geometric centre of the 
ring. It is therefore of interest to ask where the O star would have been
at the time of formation of the HII region. In the absence of proper
motion of the ring itself, a backwards extrapolation of the O star
proper motion would have placed the latter within the wedge-shaped $1\sigma$
limits delineated in blue in Fig. \ref{fig6}. Closest approach would
have occurred $\sim3.2\times10^4$ yr ago, with the O star
missing the geometric centre by approximately $2.4\sigma$.

However, the molecular ring undoubtedly has some proper motion of its own, 
and this would need to be taken into account in order to determine the 
previous trajectory of the O star relative to the ring. Unfortunately 
there are currently no observational data that would permit such a 
determination.  For example, even though several compact sources are present in
the molecular ring \citep{figueira2017}, none has a convincing match
with a {\it Gaia\/} object. Conversely, although matches exist between
{\it Gaia\/} objects and published positions of YSOs \citep{zav07,martins2010},
none has a convincing association with structure in the molecular ring.
So the only available information on the transverse proper motion of the
ring is the mean Galactic rotation in the vicinity, 
namely -0.84 mas yr$^{-1}$ in longitude (equivalent to -5.4 km s$^{-1}$), based
on the most recent Galactic rotation curve \citep{russ17}. The corresponding
radial velocity would be -8.5 km s$^{-1}$, consistent with the 
observed -8 km s$^{-1}$ radial velocity of the molecular ring\footnote{These 
two radial
velocities would not be independent if we simply used the kinematic distance
(1.35 kpc), but there exists an independent photometric estimate of 1.33 kpc 
\citep{zav07}.} \citep{tor15,and15,san18}.

So in summary, in the absence of any measurements of proper
motion of the molecular ring, we cannot establish the previous trajectory of
the O star with respect to it. However, if we make the assumption that 
the O star was, at some previous time, located in the centre of the ring,
then its current displacement can be understood as a natural consequence
of proper motion since that time.
We now explore the implications of that assumption, starting with the
epoch of formation of the O star, whose mass is assumed to be $\sim20\,M_\odot$.
For simplicity we assume that its ionising radiation is isotropic and that
the ambient ISM is approximately uniform.

The O star would have evolved rapidly, with 
the photospheric temperature increasing on a timescale $\sim10^4$ yr 
on the approach to the main sequence \citep{hos09,dav11}.
It would then have ionised the surrounding gas on an even shorter timescale
($\sim40$ years), at which point the resulting spherical HII region would have
expanded and swept up a dense surrounding shell in a fashion similar
to that envisaged by \citet{zav07} and \citet{deharv09}. The key difference
is that, following the formation of the shell, the O star would have continued 
to move through the cloud. The shell would continue to expand, but the
dust contained within it would
form a barrier to the ionising photons, thus maintaining a spherical
boundary to the HII region. The observed bubble would then represent
the fossilised remnant of the initial expansion of the HII region.

The time, $t$, taken for the HII region to expand from the radius, $R_{\rm si}$,
of the initial Str\"omgren sphere to its current radius, $R$ (1.7 pc),
can be obtained from the analytical expressions given by \citet{spit78}, 
which yield
\begin{equation}
t = \frac{4}{7}\frac{R_{\rm si}}{c_{\rm s}} \left[\left(\frac{R}{R_{\rm si}}
\right)^\frac{7}{4} - 1\right],
\end{equation}
where $c_{\rm s}$ is the sound speed in ionised hydrogen and 
$R_{\rm si}$ is given by
\begin{equation}
R_{\rm si} = \left[
\frac{3}{4\pi} \frac{\cal L} { n_{\rm i}^2 \alpha_{_{\rm B}}}
\right]^\frac{1}{3}.
\end{equation}
\noindent Here $n_{\rm i}$ is the number density of neutral hydrogen 
atoms in the
ambient cloud, ${\cal L}$ is the ionising photon luminosity, and
$\alpha_{_{\rm B}}$ is the case-B recombination coefficient for 
ionised hydrogen at $10^4$ K, which we take as 
$2.7\times10^{-13}$ cm$^3$ s$^{-1}$ \citep{bis15}. Further, assuming
$n_{\rm i}$ to be in the range 1000-3000 cm$^{-3}$ and ${\cal L}=10^{48.04}$ 
photons s$^{-1}$
\citep{zav07,art11} together with a sound speed, $c_{\rm s}=13$ km s$^{-1}$, 
appropriate to ionised hydrogen at $10^4$ K, we
obtain a range of expansion times of 0.23--0.42 Myr, consistent with previous
estimates \citep{zav07,and15,mack15}.

During this interval, the O star will have moved a projected distance
$\Delta r$ from the location of ionisation onset to its current location.
Based on the geometry illustrated in Fig. \ref{fig6}, we estimate this distance
to be 0.93 pc. The component of O star velocity with respect to
the shell in a plane transverse to the line of sight is then 
$v_{_{\rm T}}=\Delta r/t$, giving a range of possible values
2.2--3.9 km s$^{-1}$. The motion would be at a position angle of
approximately 241$^\circ$ with respect to Galactic north. Doing a vector
subtraction of these velocities from the observed proper motion of the
O star (from {\it Gaia\/}) gives the implied proper motion of the molecular 
ring, whose components in the longitude and latitude directions would then 
be $\sim-25$ km s$^{-1}$ and $\sim5$ km s$^{-1}$, respectively.

We note that our explanation for the spherical shell, in terms of a
fossilised remnant of initial HII region expansion, is not consistent with the 
results of the
simulation by \citet{mack15} which predict an egg-shaped boundary for the
HII region. Those simulations were nevertheless based on reasonable parameters 
for RCW 120, which included an O star moving with velocity 4 km s$^{-1}$
through a uniform ISM of hydrogen number density 3000 cm$^{-3}$, taking account
of the effects of a stellar wind bubble. They predict that after an elapsed 
time of 0.4 Myr, the ratio of major to minor axes of the ionisation boundary
would be approximately 1.4. This, however, is inconsistent with the 
observational result shown in Fig. \ref{fig6} which
indicates no more than $\pm5$\% deviation from circularity.
A possible explanation for the discrepancy is the neglect of dust
in the \citet{mack15} simulations, since dust may have a significant effect
on the shape of the ionised region. The importance of dust is suggested
by the fact that more than 50\% of all of the ionising photons are absorbed 
by dust within the HII region in the initial stages of evolution \citep{art11}.
In the context of the evolutionary scenario outlined above, the dust might 
be expected to prevent ionising
photons from travelling beyond the boundaries of the expanding shell (whose
geometry is established early on), thus
preventing the ionised region from morphing into an egg shape. 
Confirmation, however, will need to await numerical simulation.

If the above model is correct, then the O star will have been born
in relative motion with its parent cloud, the transverse component of
its relative velocity being $\sim2$--4 km s$^{-1}$. The question then
arises as to the origin of this motion. Possibilities are:
\begin{enumerate}
  \item The O star resulted from a cloud-cloud collision \citep{habe92,tor15}.
  \item The O star was born in a single cloud, acquiring its space velocity
from supersonic turbulence. \citet{mack15} have postulated that space
velocities $\sim2$--5 km s$^{-1}$ can arise from such a mechanism, but
no quantitative predictions appear to be available, based either on
simulations or theory. It is therefore unclear whether it can account
for stellar motion in RCW 120.
  \item Dynamical ejection from a binary or higher-order multiple system
\citep{renz18}. Such a mechanism could, in principle, provide a
massive star with a space velocity of a few km s$^{-1}$ with respect to
its parent cloud, but it is probably not viable in the case of RCW 120 since
the O star has no nearby companions of comparable mass.
\end{enumerate}

So based on currently available information, the cloud-cloud 
collision model \citep{tor15} seems to provide the most viable scenario.
The formation of the O star is then 
attributed to the collision of two clouds with an impact velocity
of 20 km s$^{-1}$ in the radial direction. This model predicts that 
the O star would continue moving through the HII region
with somewhat less than the original impact speed.

An interesting feature of the dust maps in Fig. \ref{fig5} is an arc-shaped
region which appears in blue in the middle panel, representing the warmest
dust detected by {\it Herschel\/} ($>36$ K), and in red in the corresponding
{\it Spitzer\/} image, representing 24 $\mu$m emission. Although its spatial
relationship with the O star is reminiscent of a bow shock, such
an interpretation is unlikely given the star's probable subsonic motion.
We base the latter on a $\sim7$ km s$^{-1}$ upper limit for the star's speed
relative to the shell, dictated by the requirement that the star still be
located within the boundary of the shell, given the inferred lower limit
(0.23 Myr) of the expansion timescale. More viable interpretations
of the arc-shaped feature include the effects of stellar
winds from subsonically-moving stars \citep{mack15}
and ``dust waves" due to the dragging 
effect of radiation pressure on flowing gas \citep{och14}.

Finally, we note that if the cloud-cloud collision model is applicable
to RCW 120, the O star would presumably be the largest member of a
whole cluster of stars produced during the collision, since
recent simulations \citep{bal15,bal17} have shown that such an event
can result in a distribution
of protostars in a web-like or hub-and-spoke configuration.
However, the latter simulations were based on a regime of
parameter space very different from that in the \citet{tor15}
model, in that they involved the collision of clouds of {\it similar\/} mass 
at {\it low\/} relative velocity ($\leq4$ km s$^{-1}$). 
Because of the likelihood that
cloud-cloud collisions account for a substantial fraction of massive star
formation it will be important to expand the scope of future simulations and
continue the analysis of observations of other objects similar to RCW 120.
\smallskip

In summary, we present evidence that the warm dust in the PDR associated
with the HII region is distributed in a spherical shell. The displacement
of the O star from the geometric centre can be explained by proper 
motion since the formation of the shell. The velocities involved are
consistent with the cloud-cloud collision model as envisaged by
\citet{tor15}, but we interpret the bubble boundary as the fossilised remnant 
of a swept-up shell rather than as the interior of a U-shaped cavity.

\section*{Acknowledgements}

We thank the referee for helpful comments.  We also gratefully acknowledge 
the support of a consolidated grant (ST/K00926/1) 
from the UK Science and Technology Funding Council. This work was performed 
using the computational facilities of the Advanced Research Computing at 
Cardiff (ARCCA) Division, Cardiff University. 
It has made use of data from the European Space Agency (ESA)
mission {\it Gaia\/} (https://www.cosmos.esa.int/gaia), processed by
the {\it Gaia\/} Data Processing and Analysis Consortium (DPAC,
(https://www.cosmos.esa.int/web/gaia/dpac/consortium). Funding
for the DPAC has been provided by national institutions, in particular
the institutions participating in the {\it Gaia} Multilateral Agreement.
The work is based, in part, on observations made with the Spitzer Space 
Telescope, which is operated by the Jet Propulsion Laboratory, California 
Institute of Technology under a contract with NASA.

\label{lastpage}

\begin{thebibliography}{99}
\bibitem[\protect\citeauthoryear{Anderson et al.}{2010}]{and10} 
  Anderson, L. D., Zavagno, A., Rod\'on, J. A. et al. 2010, A\&A, 518, L99

\bibitem[\protect\citeauthoryear{Anderson et al.}{2015}]{and15} 
  Anderson, L. D., Deharveng, L., Zavagno, A. et al. 2015, ApJ, 800, 101

\bibitem[\protect\citeauthoryear{Arthur et al.}{2011}]{art11} 
  Arthur, S. J., Henney, W. J., Mellema, G. et al. 2011, MNRAS, 414, 1747

\bibitem[\protect\citeauthoryear{Balfour et al.}{2015}]{bal15} 
  Balfour, S. K., Whitworth, A. P., \& Hubber, D. A., \& Jaffa, S. E.  2015, 
  MNRAS, 453, 2471

\bibitem[\protect\citeauthoryear{Balfour et al.}{2017}]{bal17} 
  Balfour, S. K., Whitworth, A. P., \& Hubber, D. A. 2017, MNRAS, 465, 3483

\bibitem[\protect\citeauthoryear{Bisbas et al.}{2015}]{bis15} 
  Bisbas, T. G., Haworth, T. J., Williams, R. J. R. et al. 2015, MNRAS,
  453, 1324

\bibitem[\protect\citeauthoryear{Davies et al.}{2011}]{dav11}
  Davies, B., Hoare, M. G., Lumsden, S. L. et al. 2011, MNRAS, 416, 972

\bibitem[\protect\citeauthoryear{Deharveng et al.}{2009}]{deharv09} 
  Deharveng, L., Zavagno, A., Schuller, F. et al. 2009, A\&A, 496, 177

\bibitem[\protect\citeauthoryear{Elia et al.}{2013}]{elia13}
  Elia, D., Molinari, S., Fukui, Y. et al. 2013, ApJ, 772, 45

\bibitem[\protect\citeauthoryear{Figueira et al.}{2017}]{figueira2017}
  Figueira, M., Zavagno, A., Deharveng, L. et al. 2017, A\&A, 600, 93

\bibitem[\protect\citeauthoryear{Griffin et al.}{2013}]{griffin13}
  Griffin, M. J., North, C. E., Amaral-Rogers, A. et al. 2013, MNRAS, 434, 992

\bibitem[\protect\citeauthoryear{Habe \& Ohta}{1992}]{habe92}
  Habe, A. \& Ohta, K. 1992, PASJ, 44, 203

\bibitem[\protect\citeauthoryear{Hildebrand}{1983}]{hil83}
  Hildebrand, R. H. 1983, QJRAS, 24, 267

\bibitem[\protect\citeauthoryear{Hosokawa \& Omukai}{2009}]{hos09}
  Hosokawa, T., Omukai, K. 2009, ApJ, 691, 823

\bibitem[\protect\citeauthoryear{Mackey et al.}{2015}]{mack15}
  Mackey, J., Gvaramadze, V. V., Mohamed, S., \& Langer, N. 2015,
  A\&A, A10

\bibitem[\protect\citeauthoryear{Marsh et al.}{2015}]{mar15} 
  Marsh, K. A., Whitworth, A. P. \& Lomax, O., 2015, MNRAS, 454, 4282

\bibitem[\protect\citeauthoryear{Marsh et al.}{2018}]{mar18} 
  Marsh, K. A., Whitworth, A. P., Smith, M. W. L. et al. 2018, MNRAS, 480, 3052

\bibitem[\protect\citeauthoryear{Martins et al.}{2010}]{martins2010} 
  Martins, F., Pomar\`es, M., Deharveng, L. et al. 2010, A\&A, 510, A32.

\bibitem[\protect\citeauthoryear{Molinari et al.}{2010}]{mol2010}
  Molinari, S., Swinyard, B., Bally, J., et al. 2010, PASP, 122, 314

\bibitem[\protect\citeauthoryear{Ochsendorf et al.}{2014}]{och14}
  Ochsendorf, B. B., Verdolini, S., Cox, N. L. J. et al. 2014, A\&A, 566, A75

\bibitem[\protect\citeauthoryear{Pavlyuchenkov et al.}{2011}]{pav13}
  Pavlyuchenkov, Ya. N., Kirsanova, M. S. \& Wiebe, D. S. 2013,
  Astronomy Reports, 57, 573

\bibitem[\protect\citeauthoryear{Poglitsch et al.}{2010}]{pog10} Poglitsch, A.,
  Waelkens, C., Geis, N., et al. 2010, A\&A, 518, L2

\bibitem[\protect\citeauthoryear{Renzo et al.}{2018}]{renz18} Renzo, M.,
  Zapartas, E., de Mink, S. E. et al. 2018, arXiv:1804.09164v1

\bibitem[\protect\citeauthoryear{Rodmann et al.}{2006}]{rod06} Rodmann, J.,
  Henning, Th., Chandler, C. J. et al. 2006 A\&A, 446, 211

\bibitem[\protect\citeauthoryear{Russeil et al.}{2017}]{russ17} Russeil, D.,
  Zavagno, A., M\`ege, P., et al. 2017, A\&A, L5

\bibitem[\protect\citeauthoryear{S\'anchez-Cruces et al.}{2018}]{san18} 
  S\'anchez-Cruces, M., Castellanos-Ram\'irez, A., Rosado, M. et al.
  2018, arXiv:1806.00724v1

\bibitem[\protect\citeauthoryear{Shetty et al.}{2009}]{shet09} Shetty, R.,
  Kauffmann, J., Schnee, S. \& Goodman, A. A. 2009, ApJ, 696, 676

\bibitem[\protect\citeauthoryear{Spitzer}{1978}]{spit78} 
  Spitzer, L. (ed.) 1978, Physical Processes in the Interstellar Medium
  (New York: Wiley)

\bibitem[\protect\citeauthoryear{Torii et al.}{2015}]{tor15} Torii, K.,
  Hasegawa, K., Hattori, Y. et al. 2015, ApJ, 806, 7

\bibitem[\protect\citeauthoryear{Walch et al.}{2015}]{walch15} 
  Walch, S., Whitworth, A. P., Bisbas, T. G. et al. 2015, MNRAS, 452, 2794

\bibitem[\protect\citeauthoryear{Zavagno et al.}{2007}]{zav07} 
  Zavagno, A., Pomar\`es, M., Deharveng, L. et al. 2007, A\&A, 472, 835
\end{thebibliography}
\end{document}